\documentclass[%
 reprint,
amssymb, amsmath,%
 aip,
]{revtex4-1}

\usepackage{graphicx}
\usepackage{dcolumn}
\usepackage{bm}

\usepackage[colorlinks=true,linkcolor=blue]{hyperref}%

\newcommand{\PC}{Physica C }
\newcommand{\IEEEas}{IEEE Trans. Appl. Supercond. }
\newcommand{\JSNM}{J. Supercond. Nov. Magn. }
\newcommand{\PRB}{Phys. Rev. B }
\newcommand{\PRL}{Phys. Rev. Lett. }
\newcommand{\APL}{Appl. Phys. Lett. }
\newcommand{\JAP}{J. Appl. Phys. }
\newcommand{\SUST}{Supercond. Sci. Technol. }
\newcommand{\NatMat}{Nat. Mater. }
\newcommand{\RMP}{Rev. Mod. Phys. }

\begin{document}
\title{Anisotropy and directional pinning in {YBa$_2$Cu$_3$O$_{7-x}$} with {BaZrO$_3$} nanorods}%
\author{N. Pompeo}
\affiliation{Dipartimento di Ingegneria and CNISM, Universit\`a Roma Tre, Via della Vasca Navale 84, 00146 Roma, Italy}%
\author{A. Augieri}
\affiliation{ENEA-Frascati, Via Enrico Fermi 45, 00044 Frascati, Roma, Italy}%
\author{K. Torokhtii}
\affiliation{Dipartimento di Ingegneria and CNISM, Universit\`a Roma Tre, Via della Vasca Navale 84, 00146 Roma, Italy}%
\author{V. Galluzzi}
\affiliation{ENEA-Frascati, Via Enrico Fermi 45, 00044 Frascati, Roma, Italy}%
\author{G. Celentano}
\affiliation{ENEA-Frascati, Via Enrico Fermi 45, 00044 Frascati, Roma, Italy}%
\author{E.Silva}
\affiliation{Dipartimento di Ingegneria and CNISM, Universit\`a Roma Tre, Via della Vasca Navale 84, 00146 Roma, Italy}%

\begin{abstract}
Measurements of anisotropic transport properties (dc and high-frequency regime) of driven vortex matter in YBa$_2$Cu$_3$O$_{7-x}$ with elongated strong-pinning sites ($c$-axis aligned, self-assembled BaZrO$_3$ nanorods) are used to demonstrate that the effective-mass angular scaling takes place only in intrinsic physical quantities (flux-flow resistivity), and not in pinning-related Labusch parameter and critical currents. Comparison of the dynamics at different time scales shows evidence for a transition of the vortex matter toward a Mott phase, driven by the presence of nanorods. The strong pinning in dc arises partially from a dynamic effect.
\end{abstract}


\date{today}

\maketitle

%
The role of nanoinclusions in the vortex pinning properties of YBa$_2$Cu$_3$O$_{7-x}$ (YBCO) films is a matter of strong interest\cite{maiorov,obradors,mcmanus,pompeoAPL07} in view of the great potential of YBCO films and tapes with strong pinning (e.g., in coated conductors). Nanoinclusions due to BaZrO$_3$ (BZO) received a particular attention.\cite{maiorov,obradors,mcmanus,pompeoAPL07,paturi,galluzziIEEE07,pompeoJAP09,mikheenkoSUST10} While in films grown by Pulsed Laser Deposition (PLD) BZO self-assembles in the shape of nanorods\cite{maiorov} of typical size $\sim 5$ nm in diameter and $30-150$ nm in length, oriented approximately along the $c$-axis (perpendicular to the film plane), chemical methods tend to yield nanoparticles instead,\cite{obradors} of typical size of $\sim 15$ nm. Such nanostructures, of typical dimension of the vortex core, are thus ideal candidates for strong core pinning: critical current density $J_c$, pinning force $F_p=J_cB$, pinning constant or Labusch parameter $k_p$ are largely enhanced by addition of BZO nanoinclusions.\cite{maiorov,obradors,mcmanus,pompeoAPL07,paturi,galluzziIEEE07,pompeoJAP09} Despite the significant technological results, the underlying vortex physics is still under debate.

The competitive effects in the vortex dynamics between the structural, unavoidable anisotropy of the YBCO matrix and the preferred orientation introduced by BZO nanorods is studied in this Letter. With respect to dc 
studies,\cite{maiorov,zuevAPL08,augieriJAP10,zuevSUST12} here we exploit the dynamics at different time scales and we show that the anisotropic-mass angular scaling exists and is limited to intrinsic properties, such as the flux-flow resistivity $\rho_{\it ff}$. By contrast, directional pinning arising from the nanorods and the layered dominates and completely destroys the angular scaling in pinning-related quantities. Finally, we find evidence for a Mott-insulator-like behavior (dynamic effect) of the vortex matter when the magnetic field is within $\sim 30^{\circ}$ with the nanorods direction.

A set of YBCO films, 120 nm thick, $c$-axis oriented, were grown on SrTiO$_3$ substrates under identical conditions by PLD from targets with BZO powders at 5\% mol. as extensively reported elsewhere.\cite{galluzziIEEE07} Transverse TEM images showed columnar-like defects, approximately perpendicular to the film plane,\cite{augieriJAP10} absent in pure YBCO films. The density of columns is consistent with an equivalent matching field of about 0.85 T. $T_c\simeq 90~$K (zero dc resistance criterion) was consistently found in all samples, with zero-field $J_c\simeq$ 3.7 MA cm$^{-2}$ at $T=$ 77 K.

Very different vortex dynamics were studied by the dc transport critical current density $J_c(H,\theta)$, and the high-frequency (48 GHz) ac transport measurements of the field-increase of the complex resistivity $\Delta\tilde\rho(H,\theta)=\tilde\rho(H,\theta)-\tilde\rho(H=0)=\Delta\rho_1(H,\theta)+\mathrm{i}\Delta\rho_2(H,\theta)$. Here, $\theta$ is the angle between the applied field $H$ and the $c$-axis.

\begin{figure}[ht]
\centerline{\includegraphics{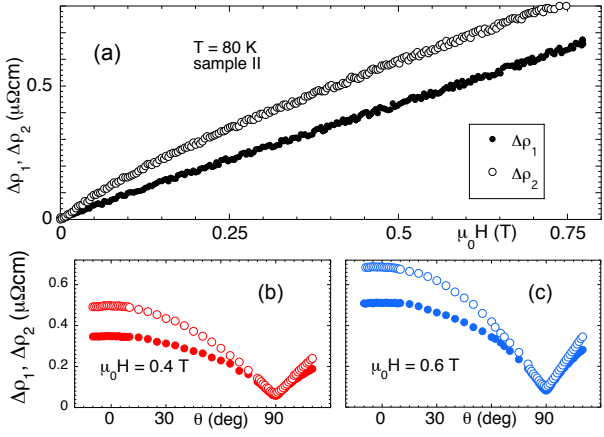}}
  \caption{Microwave complex resistivity data. Full circles: real part; open circles: imaginary part. (a): field-sweep at $\theta=0^{\circ}$. (b): angular rotations at $\mu_0H_1=0.4~$T and (c) $\mu_0H_2=0.6~$T.}
\label{muwraw}
\end{figure}

Microwave measurements were performed on unpatterned samples I and II by a sapphire cylindrical dielectric resonator operating at $\sim$48 GHz\cite{pompeoJS07} and modified to work in transmission. Special care was taken\cite{stratiSTOnoi} in order to avoid the well-known substrate resonances of SrTiO$_3$. Field sweeps ($\mu_0 H\leq$0.8 T) were performed at different angles, as well as angular rotations at fixed fields. Due to the small signal when $\theta\rightarrow 90^\circ$, we performed measurements at $T\simeq 80$ K, sufficiently below $T_c$ to avoid contributions to the microwave response from pairbreaking, but still with an acceptable signal-to-noise ratio. Fig.\ref{muwraw}a reports a typical field sweep $\Delta\tilde\rho(H,\theta=0^{\circ})$, Fig.s \ref{muwraw}b and \ref{muwraw}c report angle-rotations at $\mu_0H_1=0.4$ T and $\mu_0H_2=0.6$ T.
Measurements of the transport critical current density $J_c(H,\theta)$ (1 $\mu$V cm$^{-1}$ criterion) were taken on sample II after patterning as a strip 30 $\mu$m wide and 1 mm long, in the standard four-contact configuration.

In YBCO with BZO nanorods we expect three sources of vortex dynamics anisotropy: 
(a) the mass anisotropy leads to larger upper critical field $H_{c2}$ as $H$ is tilted away from the $c$-axis: at fixed field we expect a reduction of the dissipation solely due to smaller reduced field $H/H_{c2}(\theta)$. No pinning is conceptually involved in this effect. 
(b) the naturally arising intrinsic pinning potentiual due to the layered structure of YBCO, and
(c) the strong pinning by BZO nanorods, the topic of major interest here.

The unambiguous identification of the directionality of pinning requires a reliable determination of the electronic mass anisotropy involved in (a). This is usually done by exploiting the so-called scaling property.\cite{blatterPRL,haoclem} Let $H_{c2}(\theta)=H_{c2}(0^{\circ})/\epsilon(\theta)$ be the angle-dependent upper critical field, where $\epsilon^2(\theta)=\gamma^{-2}\sin^2\theta+\cos^2\theta$ in continuous (3D) anisotropic superconductors, and $\gamma=H_{c2}(90^{\circ})/H_{c2}(0^{\circ})\simeq 5\div 8$ is the YBCO anisotropy ratio. In the London approximation a thermodynamic or intrinsic transport property $q$ depends on the applied field $H$ and angle $\theta$ as $q(H,\theta)=s_q q[H\epsilon(\theta)]$, where $s_q$ is a scale factor typically equal to 1, $\gamma^{-1}$ or $\epsilon^{\pm1}(\theta)$ depending on the observable to be scaled (e.g., $s_\rho=1$ for in-plane resistivity). The typical procedure is to look for an experimentally determined angular function $f(\theta)$ such that the data for all angles and fields collapse over a single curve when plotted against the reduced field $H/f(\theta)$, and then compare $f$ to the theoretical expectation.\cite{silvaPRB97}

The scaling rule is {\em not} theoretically grounded when sources of anisotropy other than the electronic mass affect the measured observable (e.g.: \textit{point} defects are likely to leave the scaling properties untouched, differently from extended defects). In fact, we found no angular scaling neither for $\Delta\tilde\rho(H,\theta)$ nor for $J_c(H,\theta)$:\cite{pompeoJSNM2012} $J_c$ is inherently not an intrinsic quantity, and $\Delta\tilde\rho$ includes contributions from free motion of vortices (which {\em is} an intrinsic phenomenon) and from pinning (extrinsic). We gain more information by extracting the genuine flux flow resistivity $\rho_{\it ff}(H,\theta)$ and the Labusch parameter $k_p$ (pinning constant) from the measured $\Delta\tilde\rho$.\cite{pompeoPRB08} This is a powerful feature of high-frequency measurements: both $\rho_{ff}$ and $k_p$ can be obtained from the measurements.

\begin{figure}[ht]
\centerline{\includegraphics{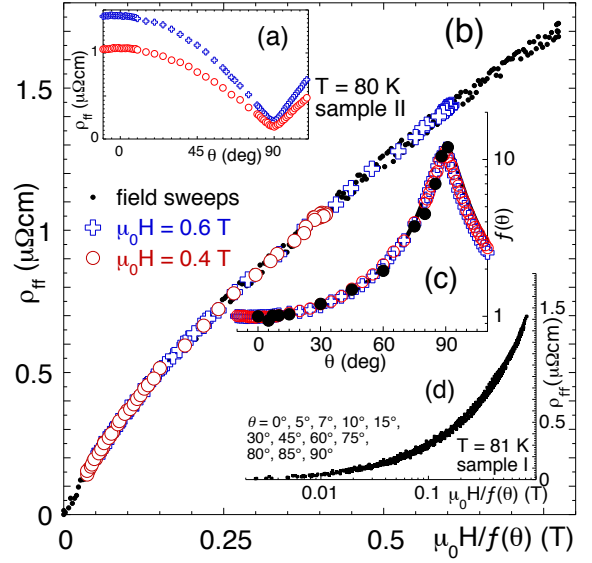}}
  \caption{Flux-flow resistivity derived from $\Delta\tilde\rho$. (a): angular dependence, sample II; (b): scaling of the data over the curve $\rho_{\it ff}(H,\theta=0^{\circ})$ with the rescaled field $H/f(\theta)$; (c): experimental $f(\theta)$ (symbols) and comparison with the theoretical 3D expression, Eq.\eqref{eq:f} (continuous line); (d) scaling for the field-sweeps at various angles in sample I. The log scale emphasizes low-fields scaling.}
\label{muwscal}
\end{figure}

The angular dependence of $\rho_{\it ff}(H_i,\theta)$ ($i=1,2$) in sample II is reported in Fig.\ref{muwscal}a. Fig.\ref{muwscal}b shows the scaling of the same data over the curve taken at $\theta=0^{\circ}$, when plotted as a function of the applied field scaled by an experimentally-determined scaling function $f(\theta)$ (we discuss $f(\theta)$ later). For completeness, Fig.\ref{muwscal}d reports the similarly scaled field-sweeps of $\rho_{\it ff}$ in sample I,\cite{pompeoPhC11} performed at several $\theta$. The log scale emphasizes the scaling at low fields. A nearly perfect scaling is obtained, both with field-sweeps and with angle-rotations.
The experimentally determined angular scaling functions for field-sweeps, $H_1$ rotations and $H_2$ rotations are reported in Fig.\ref{muwscal}c, and clearly describe a unique curve that thus does not depend on the sample and on the applied field, consistently with an intrinsic phenomenon. This is the first result of this Letter: the genuine $\rho_{\it ff}$ obeys the scaling rule, as opposed to $J_c$ and to the raw $\Delta\tilde\rho$.

In order to compare $f(\theta)$ with theoretical expressions one must take into account that the Lorentz force changes with $\theta$, due to the circular microwave currents in the experimental setup\cite{pompeoJS07}. Using a detailed theory of such cases,\cite{pompeoNote} and the experimental behavior of our data, $\rho_{\it ff}(H)\propto H^{\beta}$ with $\beta\simeq 0.8$, our measured flux flow resistivity is still expected to scale: $\rho_{\it ff} (H,\theta)=\rho_{\it ff}(H/f(\theta))$, with the expected $f(\theta)\neq \epsilon^{-1}(\theta)$ due to the varying Lorentz force contribution. We get:\cite{pompeoNote}
\begin{equation}
\label{eq:f}
    f(\theta)=\epsilon^{-1}(\theta)
    \times
    \left[\frac{\gamma^{-2}\sin^2\theta+\cos^2\theta}{\frac{\gamma^{-2}}{2}\sin^2\theta+\cos^2\theta}\right]^{1/\beta}
\end{equation}
where the term in square brackets is the Lorentz force correction. Eq.\eqref{eq:f} has no fit parameters: the experimental value at $\theta=90^{\circ}$ fixes $\gamma=5$ 
(note that $\gamma$ appears also in $\epsilon(\theta)$, and $\epsilon(90^{\circ})=\gamma^{-1}$), 
within the range of accepted values for YBCO. As shown in Fig.\ref{muwscal}c, Eq.\eqref{eq:f} is in excellent agreement with the data.
We stress that the effects of the mass anisotropy and of the Lorentz force have been obtained from the analysis of $\rho_{\it ff}$ alone, and they are not subjected to further adjustments. 

We now turn to the issue of the directionality of pinning, using the data taken with increased angular accuracy in sample II. To obtain the angular dependence of $k_p(H,\theta)$ from the data, one has to remove the Lorentz force angular contribution, which enters in the measured quantity analogously to what happens for $\rho_{\it ff}$. Since from Eq.\eqref{eq:f} one can explicitly extract such correction, the true $k_p(H,\theta)$ can be obtained from the data.\cite{pompeoPRB08,pompeoNote}

\begin{figure}[ht]
\centerline{\includegraphics{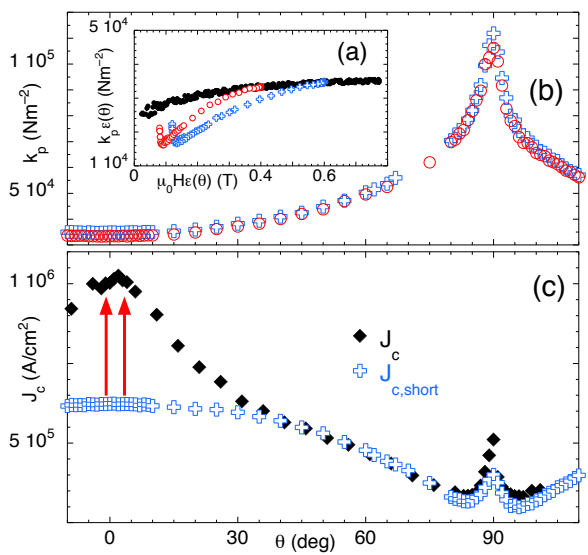}}
  \caption{(a) Failure of the BGL scaling (Ref. \onlinecite{blatterPRL}) of $k_p(H_i,\theta)$ (open symbols) over the curve $k_p(H,\theta=0^{\circ})$ (full circles). (b): angular dependence of the pinning constant $k_p$, same fields and symbols as in Fig.\ref{muwscal}b.  (c) Comparison between $J_{c,short}(0.6\;{\rm T},\theta)$ (see Eq.\eqref{eq:jcmw}) at  $T=~$80 K and $J_c(1\;{\rm T},\theta)$ at 77 K. Red arrows demonstrate the Mott-insulator effect.}
\label{pin}
\end{figure}

In Fig.\ref{pin}b we plot $k_p(H_i,\theta)~(i=1,2$): a large peak can be seen at $\theta= 90^{\circ}$, where both the mass anisotropy and the $a-b$ plane pinning are expected to decrease the dissipation. No directional effect acting around $\theta\sim 0^{\circ}$ seems evident. These observations do not imply that the anisotropy of $k_p(\theta)$ is dictated by the mass anisotropy: for this to be true, the so-called BGL scaling\cite{blatterPRL} should apply, and $s_{k_p}^{-1}(\theta)k_p(H,\theta)$ vs.  $H\epsilon(\theta)$ would describe a single curve. Since $s_{k_p}^{-1}(\theta)= \epsilon(\theta)$,\cite{blatterPRL,pompeoNote} we plot in Fig.\ref{pin}a the data of Fig.\ref{pin}b as $k_p(H,\theta)\epsilon(\theta)$ vs. $H\epsilon(\theta)$, together with $k_p(H,0^{\circ})$ as derived from the field sweep with $H\parallel c$: the scaling clearly fails, the main origin of the anisotropy of $k_p$ is {\em not} the mass anisotropy. The angular dependence in Fig.\ref{pin}a necessarily originates from other effects, demonstrating the existence of directional pinning unrelated to the anisotropic mass, stronger when the field is aligned with the nanocolumns (large $H\epsilon(\theta)$ in Fig.\ref{pin}a) or with the $a$-$b$ planes ($\theta\rightarrow90^\circ$ and small $\epsilon(\theta)$ in Fig.\ref{pin}a).
The comparison of $k_p$, as obtained from microwave measurements, and the dc critical current density $J_c$, elucidates further the physics of vortex matter in YBCO with nanocolumns. The maximum pinning force per unit length is $k_pr_p=J_c\Phi_0$ where $r_p$ is the pinning range of a defect, and by very general arguments (e.g., Ref. \cite{koshelev}), $r_p\sim\xi$ for core pinning. In inclined fields one has\cite{blatterone,tinkham} $r_p(\theta)\sim\xi_{ab}\epsilon(\theta)$, where $\xi_{ab}$ is the $a$-$b$ plane coherence length. Thus, from microwave data we define an equivalent critical current density as:
\begin{equation}
\label{eq:jcmw}
	J_{c,short}(H,\theta)=c \frac{k_p(H,\theta) \xi_{ab} \epsilon(\theta)}{\Phi_0}
\end{equation}
\noindent where $c\sim 1$, and the subscript ``short" indicates the very important point that the dc and microwave vortex dynamics differ substantially: $J_c$ is measured in the regime where flux lines are depinned by a sufficiently strong direct current, and steady motion arises. $J_{c,short}$ is defined in the {\em subcritical} microwave currents regime, where the microwave field induces only very short ($\sim 1\;$\AA) oscillations of the vortices around their equilibrium position.

Comparison of $J_{c,short}$ to the actually measured dc $J_c$ is shown in Fig.\ref{pin}c. Common features are a maximum centered at $0^\circ$, a local minimum around $85^\circ$ and a steep peak at $90^\circ$. The latter clearly originates from $a$-$b$ planes pinning whereas the former is ascribed to the correlated pinning due to BZO inclusions.\cite{augieriIEEE09} It should be noted that the maximum at $0^\circ$ is extremely wide, thus suggesting that BZO nanorods are strong pinning centers even in tilted magnetic field. Moreover, with reasonable $\xi_{ab}(T)=\xi_0/\sqrt{1-(T/T_c)^2}$, where $\xi_0=12\;$\AA, and $c\sim 0.16$, the short-range-dynamic ($J_{c,short}$) and long-range dynamic ($J_c$) curves exactly coincide in the intermediate angular range $35^\circ<\theta<80^\circ$. Thus, in the angular range where flux lines are most likely segmented between nanorods, the angular dependences of $J_c$ and $J_{c,short}$ are identical, despite the very different dynamics.
However, $J_{c,short}(\theta)$ and $J_c(\theta)$ strongly depart from each other as the field aligns with the nanorods: for $\theta\lesssim35^{\circ}$ $J_{c,short}$ shows only a broad hump, over which a second, steeper hump appears in $J_c$. The further enhancement of $J_c(\theta)$ is then a \textit{dynamic} effect: the largest peak at $\theta=0^{\circ}$ appears only when large vortex displacements are involved (dc). Thus, we can state the important result that nanocolumns enhance $J_c$ by means of two distinct mechanisms: strong core pinning, which gives rise to the broad hump in $J_c(\theta)$, and the \textit{reduced vortex mobility} effect, which gives rise to the additional peak in $J_c(\theta)$ for $\theta<35^\circ$. We argue that the latter effect is a manifestation of the predicted Mott-insulator phase for fluxons:\cite{mott} essentially, each vortex is pinned by an extended defect and when it is forced to move away from it by the external current, its mobility is drastically reduced by the lack of free sites where to move. When the field is tilted away from the nanocolumns, fluxons are no longer strongly pinned by the extended defects for the whole length, and the Mott-insulator effect disappears at a certain angle ($\sim35^\circ$ in our case). Deformation of the flux lines (steplike) may yield stronger pinning than with point pins only, hence the same wide hump in $J_{c,short}(\theta)$ and $J_c(\theta)$.

This scenario is fully consistent with the results of DC characterizations:\cite{augieriJAP10} there, we observed\cite{augieriJAP10} a change of regime in the pinning force $F_p$ approaching the equivalent matching field (the field where each linear defect accommodates one flux line). Above that field, the Mott-insulator effect vanishes and the pinned vortex matter behaves similarly to a more or less disordered lattice in presence of strong point pins. Strong extended pinning 
has been observed
 below the pseudo-matching field, as revealed by careful analysis of the magnetization relaxation,\cite{miuPRB12}
 with a crossover to point pinning at higher fields.\cite{zuevSUST12}
 \begin{figure}[ht]
\centerline{\includegraphics{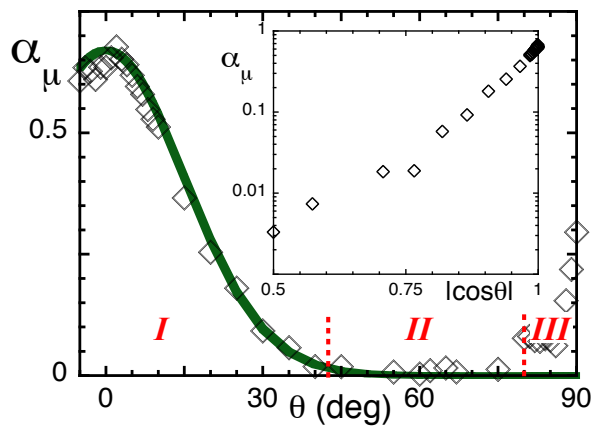}}
  \caption{Main panel: the dynamic parameter $\alpha_\mu~vs.~\theta$ (diamonds) shows three regions: \textit{I}, where dynamic effects (Mott-insulator vortex phase) increase $\alpha_\mu$; \textit{II}, where core pinning (segmented flux lines) dominates; \textit{III}, where $a$-$b$ plane pinning is important. Continuous green line: empirical fit with Eq.\eqref{eq:rpfit}. Inset: in region \textit{I} the increase in $\alpha_\mu$ depends exponentially on the component of the field along the nanocolumns.}
\label{figrp}
\end{figure}

We now estimate quantitatively the dynamic effect on pinning. A natural way to define the angular dependence of the dynamic effect is to introduce the parameter:
\begin{equation}
\label{eq:param}
	\alpha_{\mu}(\theta)=\frac{J_c(\theta)-J_{c,short}(\theta)}{J_{c,short}(\theta)}
\end{equation}
When no dynamic effect is present, that is the critical current density does not depend on the vortex displacement, $\alpha_{\mu}=0$. From the argument that led to Eq.\eqref{eq:jcmw} one can see that $\alpha_{\mu}$ can be interpreted as the percentage increase of $r_p$ due to the dynamic effect. In Fig.\ref{figrp} we plot $\alpha_{\mu}(\theta)$.  As it can be seen, $\alpha_{\mu}\neq 0$ when the field aligns with the directional pinning, it drops exponentially with $\cos\theta$, the projection of the field along the nanocolumns (inset of Fig.\ref{figrp}), and vanishes at intermediate angles, before a further increase close to the $a$-$b$ planes. The following empirical expression\cite{notealpha} describes very well $\alpha_{\mu}(\theta)$:
\begin{equation}
\label{eq:rpfit}
	\alpha_\mu(\theta)=\alpha_\mu(0)\times \frac{e^{\zeta\cos\theta}-1}{e^{\zeta}-1}
\end{equation}
where $\zeta\simeq 14.5$ is obtained from the exponential behavior for $\cos\theta > 0.9$. The physical significance of the empirical fit is that the dynamic Mott-insulator pinning effect is a fast vanishing function of the fraction of the flux line that can be pinned and ``caged" by the extended defect. 

In conclusion, by comparing the angle and field dependence of vortex dynamics over different time scales (dc critical current and microwaves) we have shown that angular scaling takes place in intrinsic quantities only, and that the pinning constant and the critical current density are dominated by directional pinning in almost the entire angular range. The overall pinning in dc is a superposition of strong core pinning and Mott-insulator dynamic effects, and we have estimated the corresponding angular ranges. We believe that our results elucidate the vortex physics in presence of elongated defects.

This work has been partially supported by the Italian FIRB project ``SURE:ARTYST'' and by EURATOM. N.P. acknowledges support from Regione Lazio.
%


\begin{thebibliography}{9}
\bibitem{notealpha} Eq.\eqref{eq:rpfit} has been constructed to have the limit $\alpha_{\mu}(\pi/2)=0$ (no effect of caging).
\end{thebibliography}
\end{document}